\documentclass[11pt]{article}
\usepackage{amsmath, amsfonts, amsthm, amssymb,verbatim}
\usepackage[mathscr]{euscript}
\usepackage[a4paper,nohead,twoside]{geometry}
\usepackage[english]{babel}
\usepackage{xspace}
\usepackage{enumitem} 
\usepackage{appendix}
\usepackage{cite}

\newtheorem{theorem}{Theorem}[section] 
\newtheorem{definition}[theorem]{Definition}
\newtheorem{lemma}[theorem]{Lemma}

\newtheorem{proposition}[theorem]{Proposition}

\numberwithin{equation}{section}
 
\usepackage[bookmarks=true,
            bookmarksopen=true,colorlinks=true,breaklinks=true,linkcolor=blue,%
            citecolor=blue]{hyperref}

\newcommand \R {\mathbb R} 
\newcommand{\scalarpr}[2]{\left<#1,#2\right>}
\newcommand{\diag}{\text{diag}\xspace}
\newcommand{\Eqref}[1]{Eq.~\eqref{#1}}
\newcommand{\Eqsref}[1]{Eqs.~\eqref{#1}}
\newcommand{\Sectionref}[1]{Section~\ref{#1}}  
\newcommand{\Defref}[1]{Definition~\ref{#1}}
\newcommand{\Lemref}[1]{Lemma~\ref{#1}}
\newcommand{\Propref}[1]{Proposition~\ref{#1}}
\newcommand{\Theoremref}[1]{Theorem~\ref{#1}}

\newcommand{\Conditionref}[1]{Condition~\ref{#1}}
\newcommand{\Conditionsref}[1]{Conditions~\ref{#1}}
\newcommand{\del}{\partial}
\newcommand{\keyword}[1]{\textbf{#1}}
\newcommand{\LPDE}[2]{\ensuremath{\widehat L(#1)[#2]}\xspace}

\newcommand{\LPDEu}[2]{\LPDE{u_0+#1}{#2}}

\newcommand{\SaHs}{\ensuremath{S^\sigma_1}\xspace}
\newcommand{\Stna}{\ensuremath{S^0}\xspace}
\newcommand{\StZna}{\ensuremath{S^0_0}\xspace}
\newcommand{\StOna}{\ensuremath{S^0_1}\xspace}
\newcommand{\St}[1]{\ensuremath{S^0(#1)}\xspace}
\newcommand{\Stu}[1]{\St{u_0+#1}}
\newcommand{\StLu}{\ensuremath{S^0_0(u_0)}\xspace}
\newcommand{\StL}{\ensuremath{S^0_0}\xspace}
\newcommand{\StH}[1]{\ensuremath{S^0_1(#1)}\xspace}
\newcommand{\StHu}[1]{\StH{u_0+#1}}
 
\newcommand{\Ssna}{\ensuremath{S^j}\xspace}
\newcommand{\SsZna}{\ensuremath{S^j_0}\xspace}
\newcommand{\SsOna}{\ensuremath{S^j_1}\xspace}
\newcommand{\Ss}[1]{\ensuremath{S^j(#1)}\xspace}

\newcommand{\SsL}{\ensuremath{S^j_0}\xspace}
\newcommand{\SsLu}{\ensuremath{S^j_0(u_0)}\xspace}
\newcommand{\SsO}[1]{\ensuremath{S^j_1(#1)}\xspace}

\newcommand{\SsH}[1]{\ensuremath{S^j_1(#1)}\xspace}
\newcommand{\SsHu}[1]{\SsH{u_0+#1}}

\newcommand{\NN}[1]{\ensuremath{N(#1)}\xspace}
\newcommand{\NNu}[1]{\NN{u_0+#1}}
\newcommand{\NL}{\ensuremath{N_{0}}\xspace}
\newcommand{\NLu}{\ensuremath{N_{0}(u_0)}\xspace}
\newcommand{\NH}[1]{\ensuremath{N_{1}(#1)}\xspace}
\newcommand{\NHu}[1]{\NH{u_0+#1}}

\newcommand{\f}[1]{\ensuremath{f(#1)}\xspace}
\newcommand{\FPDE}[2]{\ensuremath{F(#1)[#2]}\xspace}
\newcommand{\FPDEu}[1]{\FPDE{u_0}{#1}}

\newcommand{\FreduOp}{\ensuremath{\mathscr F}(u_0)\xspace}
\newcommand{\Fredu}[1]{\ensuremath{\mathscr F}(u_0)[#1]\xspace}

\newcommand{\RR}[1]{\ensuremath{\mathcal{R}[#1]}\xspace}

 \newcounter{mnotecount}[section]
 \let\oldmarginpar\marginpar
 \renewcommand\marginpar[1]{\-\oldmarginpar[\raggedleft\footnotesize #1]%
 {\raggedright\footnotesize #1}}

\begin{document}

\title{Quasilinear symmetric hyperbolic Fuchsian systems
\\
in several space dimensions}

\author{Ellery Ames\footnote{Department of Physics, University of
    Oregon, Eugene, OR 97403, USA. Research partially supported by NSF
    grant  PHY-0968612.} \hskip0.cm, Florian Beyer\footnote{Department
  of Mathematics and Statistics, University of Otago, P.O. Box 56, Dunedin 9054, New Zealand.} \hskip0.cm, James Isenberg\footnote{Department of Mathematics, University of Oregon, Eugene, OR 97403, USA. Research partially supported by NSF grant  PHY-0968612.}
\\
and Philippe G. LeFloch\footnote{Laboratoire Jacques-Louis Lions \& Centre national de la recherche scientifique,
Universit\'e Pierre et Marie Curie (Paris 6), 4 Place Jussieu, 75252 Paris, France. Email: contact@philippelefloch.org. Research partially supported by ANR grant SIMI-1-003-01.} 
} 

\maketitle

\begin{abstract}
We establish  existence and uniqueness results for the singular initial value problem associated with a class of quasilinear, symmetric hyperbolic, partial differential equations of Fuchsian type in several space dimensions. This is an extension of  earlier work by the authors for the same problem in one space dimension. 
\end{abstract}


\section{Introduction}
\label{sec:introduction}

In previous work \cite{Ames:yV5l9m6A}, we have established existence and uniqueness results for the singular initial value problem associated with quasilinear systems of partial differential equations (PDE) of Fuchsian type, with non-analytic coefficients and in one space dimension. In the present paper, we generalize these results to systems in several space dimensions. While much of the earlier work with Fuchsian systems involves PDE systems with analytic coefficients, here (and in \cite{Ames:yV5l9m6A}) we allow coefficients which are $C^{\infty}$, are of finite differentiability, or are elements of certain weighted Sobolev spaces. We recall that Fuchsian techniques have found many applications in the study of nonlinear wave equations
\cite{Kichenassamy:1996wy}, in mathematical physics
\cite{Kichenassamy:2007tr}, and in 
mathematical cosmology and the study of Einstein's field equations of general relativity  \cite{Kichenassamy:2007tr,Rendall:2000ki,Kichenassamy:1999kg,Beyer:2010fo,Beyer:2010tb,Beyer:2010wc,Ames:yV5l9m6A,Andersson:2001fa,Stahl:2002bv,Isenberg:2002ku,ChoquetBruhat:2004ix,ChoquetBruhat:2006fc,Beyer:2011uz,Beyer:2011ce}.

The earliest applications of Fuchsian techniques to mathematical cosmology concerned equations with
analytic coefficients. One such theorem by Kichenassamy and Rendall
\cite{Kichenassamy:1999kg} has been applied to several families of cosmological solutions of Einstein's equations. Existence theorems in the class of smooth (but not analytic) functions 
have proven more difficult to establish. Rendall in \cite{Rendall:2000ki} relies on energy techniques for symmetric hyperbolic systems 
and establishes that a sequence of analytic solutions converges to a smooth solution; 
he then applies his theorem to deal with the Gowdy vacuum
equations with $T^3$ spatial topology. In subsequent work, this technique has been applied by
St\r{a}hl \cite{Stahl:2002bv} with partial success in attempts  to handle the Gowdy
equations with $S^2 \times S^1$ and $S^3$ spatial topology, and has also been applied to a family of
polarized $T^2$--symmetric solutions to the Einstein equations by
Clausen \cite{Clausen:2007vq}. 

More recently, a Fuchsian method has been introduced by  Beyer and LeFloch  \cite{Beyer:2010tb,Beyer:2010fo,Beyer:2010wc,Beyer:2011ce} 
who define solutions in certain smooth and Sobolev-type function spaces  
for semilinear symmetric hyperbolic systems; they apply
their theory to establish the existence of a large class of Gowdy spacetimes with $T^3$ spatial  topology and with certain asymptotic properties in the neighborhood of the singularity. In their approach, 
a sequence of regular Cauchy problems (obtained by working away from the singularity) is introduced 
and it is shown that their solutions converge to a solution of a \textit{singular} initial value problem (defined below), provided
that suitable regularity assumptions in weighted Sobolev spaces hold. The follow--up theorems 
in \cite{Ames:yV5l9m6A} and in the present paper provide an extension of  \cite{Beyer:2010fo,Beyer:2010tb,Beyer:2010wc,Beyer:2011ce} 
to the class of quasilinear equations. Other results in the
smooth class are found in the literature and require various structural hypotheses of the Fuchsian systems under consideration and 
the space of solutions; cf.~\cite{Rendall:2000ki} for further references and, for the Sobolev class, 
\cite{Kichenassamy:2007tr,Kichenassamy:1996hr}. 
A particular distinguishing feature of our approach in \cite{Ames:yV5l9m6A,Beyer:2010fo,Beyer:2010tb}, as well as in this paper, 
is that an approximation scheme is at the core of the method. This scheme  can be implemented 
for numerical computations straightforwardly and contains useful built-in convergence and error estimates \cite{Amor:2009a,Beyer:2010tb,Beyer:2010wc}. 

As mentioned above, the present paper generalizes our results in \cite{Ames:yV5l9m6A} to an arbitrary number of space dimensions. In order to achieve this result, we work with dimension-dependent Sobolev embeddings and inequalities, and we  track all dimension-dependent arguments through the construction. Indeed, Sobolev inequalities are crucial in several places in the argument, and we refer the reader to the beginning of \Sectionref{sec:proof}, below,  for further details. 

The paper is organized as follows. In \Sectionref{sec:setupterminology}, we introduce the class of PDE systems of primary interest, referred to as quasilinear symmetric hyperbolic Fuchsian systems, and we discuss some basic terminology relevant to the singular initial value problem for such systems. We then formulate our main existence and uniqueness theorem; cf.~Theorem~\ref{theorem:maintheorem}.
We also include, as an example application, a discussion  of a nonlinear Euler-Poisson-Darboux type equation; this example serves to exhibit important features arising with the Einstein system.
 Next, \Sectionref{sec:proof} is devoted to the proof of \Theoremref{theorem:maintheorem} and, in particular, discusses some of the energy estimates needed. Concluding remarks are found in \Sectionref{sec:conclusions}.


\section{Statement of the main result}
\label{sec:setupterminology}

\subsection{Quasilinear Fuchsian systems}

We are concerned with the following system of PDEs for the unknown $u: (0, \delta] \times T^n \to \mathbb R^d$: 
\begin{align}
\label{eq:pde}
 S^0(t,x,u) Du + S^j(t,x,u) t \partial_j u + N(t,x,u) u = f(t,x,u),
\end{align}
where each of the maps $\Stna$ and $\Ssna$ is a symmetric $d \times d$ matrix-valued function of the spacetime coordinates\footnote{While all results here are stated for functions defined on the spatial manifold $T^n$, our techniques and results could be extended in similar form (with suitably adjusted function spaces) for other spatial manifolds, including $\R^n$.}
$(t,x)$ and of  the unknown $u$ (but not  of the derivatives of $u$), while
$f=f(t,x,u)$ is a prescribed $\R^d$--valued function, and 
$N=N(t,x,u)$ is a matrix-valued function of $(t,x,u)$.  
It is convenient to scale the time derivative operator and set $D := t \,  \partial_t= t \frac{\partial}{\partial t} =x^0 \frac{\partial}{\partial x^0}$, while $\partial_j := \frac{\partial}{\partial x^j}$ for $j=1,...,n$.
For notational convenience, we shall often leave out the arguments $(t,x)$, and instead use the short-hand
notation $\St{u}$, $\Ss{u}$, $\NN{u}, \f{u}$. Notationally, $\St{u}$ (for instance) may be regarded as a map $u\mapsto \St{u}$ between two function spaces (as discussed below) and, therefore, may be written as $\St{u}(t,x)$.
The notation 
\[
\LPDE{u}{v}(t,x):= S^\sigma(t,x,u) \, t \partial_\sigma v + N(t,x,u) v
\] 
is also  used throughout this paper, so that the left-hand side of \Eqref{eq:pde} can be written as $\LPDE{u}{u}$. Here, $\sigma=0,1,...n$, so that the above summation includes the time derivative term as well as the space derivative terms.

Presuming (as stated explicitly below in \Defref{def:QSHFsystem}) that the matrices $S^\sigma$ and $N$ are non-singular in a neighborhood of $t=0$, we observe that with our choice of coordinates, the PDE system \eqref{eq:pde} is singular exactly at $t=0$  in the sense that the PDE coefficients vanish there (and nowhere else nearby). 
Hence the singularities which we consider in this paper are ``spacelike''. Apart from this, the convention that the equations are singular at $t=0$ does not result in a loss of generality, since for solutions whose singularity locus is rather at $t^\prime = \psi(x)$, one could always introduce the new time coordinate $t = t^\prime - \psi(x)$.

At this point the reader may wonder why the zero--order term $\NN{u} u$ is included in the principal part and not in the source $f(u)$. We leave these terms separate since, by convention, $f$  includes terms of ``higher order" while the term $\NN{u} u$  contains terms of the same order as the principal part in $t$ (as $t \to 0$). Specific conditions on $\NN{u}$ and $f(u)$ are stated and discussed below.

In order to measure the regularity and the decay of both the solutions and the coefficients near the singularity $t=0$, we introduce a family of weighted Sobolev spaces. Letting  $\mu: T^n \to \mathbb R^d$ be a smooth\footnote{By ``smooth", we mean that sufficiently many derivatives exist and are continuous. It is straightforward to check how many derivatives are necessary in each argument and this is left to the reader.}
 function, we define the matrix 
\begin{equation}
\label{eq:rmatrix}
\RR{\mu}(t,x) := \diag\left( t^{-\mu_1(x)},...,t^{-\mu_d(x)} \right)
\end{equation}
and use it to define the following norm for functions
 $w: (0, \delta] \times T^n \mapsto \R^d$: 
\begin{align}
||w||_{\delta, \mu, q} := & \sup_{t \in (0,\delta]} ||\mathcal R[\mu] w ||_{H^q(T^n)} \\
= & \sup_{t \in (0,\delta]} \left( \sum_{|\alpha| = 0}^q \int_{T^n} |\partial^\alpha (\mathcal R[\mu] w) |^2 dx \right)^{1/2};
\nonumber
\end{align}
here $H^q(T^n)$ denotes the usual Sobolev space of order $q$ on the $n$--torus $T^n$, 
$\alpha$ denotes a partial derivative multi-index, and the standard Lebesgue measure is used for the integration. 
Note that this norm only controls spatial derivatives. Also note that 
the derivatives $\del^\alpha$ operate on both the matrix $\RR{\mu}$ and $w$; if $\mu$ is not constant, then logarithms in $t$ to the power $|\alpha|$  have to be controlled.

Next, we define the function space $X_{\delta, \mu, q}(T^n)$ to be the completion of the set of functions $w \in C^\infty\left((0,\delta]\times T^n \right)$ for which the above norm is finite. A closed ball of radius $r$ about $0$ in $X_{\delta, \mu, q}(T^n)$  is denoted by $B_{\delta, \mu, q,r}(T^n)$. Note that in our discussion below, we include the argument $T^n$ only if we wish to emphasize the dimensional dependence; we shall often write $X_{\delta, \mu, q}$ in place of $X_{\delta, \mu, q}(T^n)$, with the argument understood to be $T^n$.  
To handle functions which are infinitely differentiable and for which we control all derivatives, we also define the space $X_{\delta, \mu, \infty} := \bigcap_{q=0}^\infty X_{\delta, \mu, q}$. In the following, we refer to quantities  $\mu$ as \keyword{exponent vectors} and we always assume that these are smooth. We write $\nu>\mu$ for two exponent vectors (of the same dimension) if each component of $\nu$ is larger than the corresponding component of $\mu$ at each spatial point.

In working with $d \times d$ coefficient matrix-valued functions (such as $S^{\sigma}$), we use analogous norms and function spaces. In these cases, each component of the normed matrix is weighted by $t$ raised to the negative of the corresponding component of a smooth {\bf exponent matrix} $\zeta: T^n \to \R^{d \times d}$ (analogous to the exponent vector $\mu$ above). For convenience, we use the notation  $X_{\delta, \zeta, q}(T^n)$ to denote these function spaces.  


\subsection{The singular initial value problem}
\label{sec:fuchsianreduction}

In contrast to the Cauchy problem for \Eqref{eq:pde}, which seeks a function $u$ that satisfies  \Eqref{eq:pde}
and that equals a specified function $u_{[t_0]}: T^n \rightarrow \R^d$ at $t=t_0>0$, the singular initial value problem seeks a solution of \Eqref{eq:pde} with prescribed asymptotic behavior in a neighborhood of $t=0$. More specifically, one prescribes a ``leading order term" $u_0$, which may be either a function or a formal power series on $(0, \delta] \times T^n$, and one looks to find a solution $u$ such that $w:=u-u_0$ suitably decays relative to $u_0$ at a prescribed rate in a neighborhood of $t=0$. Substituting $u=u_0+w$ into \Eqref{eq:pde}, one obtains a PDE system for $w$, which takes the form 
\begin{equation}
\label{eq:reducedeq}
\LPDEu{w}{w} = \Fredu{w},
\end{equation}
where $ \Fredu{w} := f(u_0+w) - \LPDEu{w}{u_0}$. The problem of existence and uniqueness for the singular initial value problem is now equivalent to establishing the 
existence and uniqueness of a solution $w$ to \Eqref{eq:reducedeq} 
with a prescribed asymptotic behavior as $t\rightarrow 0$. 
Using the spaces $X_{\delta, \mu, q}$ to prescribe the asymptotic fall-off rate, we now state the singular initial value problem in the following form. 

\begin{definition}[The singular initial value problem]
\label{def:SIVP}
For a given choice of a  leading order term $u_0$ and the parameters $\delta, \mu,q$,
the \keyword{singular initial value problem} consists of finding a solution $w \in X_{\delta, \mu, q}$  to the system \Eqref{eq:reducedeq} 
and, therefore, a solution $u = u_0 + w$ to the system \Eqref{eq:pde}.   
\end{definition}

How does one choose a leading order term? One method, described in \cite{Kichenassamy:2007tr}, is to find a power series in $t$ with coefficients depending possibly on spatial variables, such that the most singular terms in the equation are canceled. Another method, which has proved useful in cosmology and in applications to the Einstein equations, is to select a leading order term as an exact solution to a {\sl simplified} system which is derived from the Einstein equations by dropping the spatial derivative terms relative to the time derivative terms. This approach leads to the so-called \emph{velocity term dominated} (VTD) system, and using the Fuchsian method we can in certain cases verify the existence of solutions with \emph{asymptotically velocity term dominated} (AVTD) behavior. We refer to  \cite{Ames:yV5l9m6A} for a general method for doing this which provides a formal series of leading order terms and
 generalizes easily to $n$ dimensions.  We make no further comments on the choice of leading order terms in the rest of this paper.


\subsection{Quasilinear symmetric hyperbolic Fuchsian systems}

We establish (cf.~\Theoremref{theorem:maintheorem}, below) that, under certain conditions, there exists a unique solution to the singular initial value problem as described in \Defref{def:SIVP}. 

\begin{definition}
\label{def:QSHFsystem}
Fix positive real numbers $\delta$ and $s$, non-negative integers $q_0$ and $q$, and an exponent vector $\mu:T^n\rightarrow\R^d$, 
together with a leading-order term $u_0 : (0, \delta] \times T^n \to \mathbb R^d$ (with so far unspecified regularity). The system \Eqref{eq:pde} is  a \keyword{quasilinear symmetric hyperbolic Fuchsian system} around $u_0$
if there exist matrices $\StLu$ that is positive definite and symmetric and independent of $t$, 
$\SsLu$ ($j=1,\ldots,n$) that are symmetric and independent of $t$, 
and $\NLu$ that is independent of $t$, all belonging to the space $H^{q_0}$, and if there exist vector functions $\beta_j : T^n \to \mathbb R^d$ ($j=1,\ldots,n$) with strictly positive components, 
such that each of the ``remainder matrices"
\begin{align*}
   \StHu{w} :=& \St{u_0 + w} - \StLu, \\
   \SsHu{w} :=& \RR{\beta_j -1}\left(\Ss{u_0 + w} - \SsLu \right),\\
   \NHu{w} :=& \NNu{w} - \NLu,
 \end{align*}
maps all elements 
$w \in B_{\delta', \mu, q,s}$ for all $\delta'\in(0,\delta]$ to elements $\StHu{w}, \SsHu{w}$ and $\NHu{w}$ in $B_{\delta', \zeta, q,r}$, where  $\zeta$ is some exponent matrix with strictly positive entries, and $r>0$ is some radius.  One also requires that $\StHu{w}$ and $\SsHu{w}$ be symmetric matrices for all $w\in B_{\delta', \mu, q,s}$.
\end{definition}

It is important clarify the notation used here and below. While the quantities $S^\alpha_0(u_0)$ and $N_0(u_0)$ are explicitly time-independent, they do depend on the $t \to 0$ behavior of the leading order term $u_0$. For convenience below, if a choice of leading order term has been fixed, we may omit the explicit dependence on $u_0$, and write simply $S_0^\alpha$ and $N_0$; the dependence on the leading order term  is then implicit. We use the same notational shorthand with $S^\alpha_1(u_0+w)$ and $N_1(u_0 + w)$, omitting the explicit dependence on $u_0$ so long as the choice of the leading order term is fixed and unambiguous. We note that $S^\alpha_1(u_0+w)$ and $N_1(u_0 + w)$ are both explicitly time-dependent, with the time dependence coming from $u_0$ as well as from $w$. We also note that so long as the system \Eqref{eq:pde} is a quasilinear symmetric hyperbolic Fuchsian system as defined above in Definition \ref{def:QSHFsystem}, the perturbing coefficients $\SaHs, N_1$ belong to the space $X_{\delta, \zeta, q}$, with $\zeta$ positive and thus are non-singular at $t =0$.  

It follows from Definition  \ref{def:QSHFsystem} that if \Eqref{eq:pde} is a 
 quasilinear symmetric hyperbolic Fuchsian system, then it is also symmetric hyperbolic for all $t \in (0, \delta]$ (so long as  $\delta$ is sufficiently small so that $S^0$ is positive definite).
Hence for smooth initial data (or for data in $H^q(T^n)$, with $q > n/2 + 1$)
prescribed at $t_0 \in (0, \delta]$, the Cauchy problem is well-posed  in the usual sense (away from $t=0$), with solutions belonging to the space  $C(I,H^q(T^n))$ for a sufficiently small interval $I \subset (0, \delta]$; see, for instance, \cite{Taylor:2011wn}. We note that since  solutions to the Cauchy problem are only  defined for $t$ bounded away from the singularity at $t=0$, we know nothing a priori  regarding  the singular behavior of these solutions, nor whether they belong to some space $X_{\delta, \mu, q}$. 

Our existence and uniqueness results rely on further structural conditions, including the following one.

\begin{definition}[Block diagonality with respect to $\mu$]  
\label{def:blockdiagonal}
Suppose that $u_0$ is a given leading-order term and $\mu$ is an exponent vector. The system (\ref{eq:pde}) is said to be  \keyword{block diagonal with respect to $\mu$} if the following commutators vanish (for all $w = u - u_0\in X_{\delta,\mu,q}$) 
\begin{align*}
  \St{u} \RR{\mu} - \RR{\mu} \St{u} = 0, \\
  \Ss{u} \RR{\mu} - \RR{\mu} \Ss{u} = 0, \\
   \NN{u}\RR{\mu} - \RR{\mu} \NN{u} = 0,
\end{align*}
where $\RR{\mu}$ is defined in \Eqref{eq:rmatrix}, and if the same property holds for all relevant spatial derivatives of $\St{u}, \Ss{u}, \NN{u}$, as well.
\end{definition}

This condition is essential in deriving our energy estimate below in Section~3.  
It ensures that the principal part operator only couples those components of the unknown which decay in $t$ at the same rate. 
Another quantity which plays a role in the derivation of energy estimates is the \keyword{energy dissipation matrix}
\begin{equation}
\label{eq:EDM}
 M_0 := \StLu\,\diag(\mu_1, ..., \mu_d) + \NLu,
\end{equation}
which is a function of the spatial coordinates $x$, only.

We can now formulate the central result of this paper, as follows. 

\begin{theorem}[Existence and uniqueness for the singular initial value problem]
\label{theorem:maintheorem}
Suppose that \Eqref{eq:pde} is a {quasilinear symmetric hyperbolic Fuchsian system} around $u_0$
(with a choice of the parameters $\delta$, $s$, $\mu$, $q$ and $q_0$ as specified in \Defref{def:QSHFsystem})
and is block diagonal with respect to $\mu$. Suppose also that $q>n/2+2$ and $q_0>n/2+1+q$.  Then, for some $\widetilde \delta \in (0, \delta]$, there exists a unique solution $u$ to \Eqref{eq:pde} whose remainder $w:=u-u_0$ belongs to $X_{\widetilde\delta, \mu, q}$ with $Dw \in X_{\widetilde \delta, \mu, q-1}$, provided the following structural conditions are satisfied:
\begin{enumerate}[label=\textit{(\roman{*})}, ref=(\roman{*})] 
\item \label{cond:EDM} The energy dissipation matrix \Eqref{eq:EDM} is positive definite at all spatial points.
\item \label{cond:source1} The map $\FreduOp : w \mapsto f(u_0+w) - \LPDEu{w}{u_0}$ is well-defined, and, for every $\delta^\prime \in (0, \delta]$, it maps 
$B_{\delta', \mu, q,s}$ to $X_{\delta', \nu, q}$ for some exponent vector $\nu > \mu$.
\item \label{cond:source2} For
all $\delta^\prime \in (0, \delta]$, for some constant $C > 0$ and for all $w, \widetilde w \in B_{\delta^\prime, \mu, q, s}$, one has  
\[|| \Fredu{w} - \Fredu{\widetilde w} ||_{\delta^\prime, \nu, q} \le C || w -\widetilde w ||_{\delta^\prime, \mu, q},\]
and, the constant $C$ below being independent of $\delta^\prime$, 
\begin{equation*}
\begin{split}
& || \Fredu{w} - \Fredu{\widetilde w} ||_{\delta^\prime, \nu, q-1} + \| \StHu{w} - \StHu{\widetilde w} \|_{\delta^\prime, \zeta, q-1} \\ 
&\quad+\sum_{j=1}^n \| \SsHu{w} - \SsHu{\widetilde w} \|_{\delta^\prime, \zeta, q-1}  + \| \NHu{w} - \NHu{\widetilde w} \|_{\delta^\prime, \zeta, q-1} 
\\
& \quad\le C || w -\widetilde w ||_{\delta^\prime, \mu, q-1}.
\end{split}
\end{equation*}
\end{enumerate}
Furthermore,  if the conditions above hold for all integers $q>n/2+2$ and $q_0>n/2+1+q$, then 
the remainder $w$ belongs to $X_{\widetilde\delta, \mu, \infty}$, while $Dw\in X_{\widetilde\delta, \mu, \infty}$. 
\end{theorem}

Analogous statements can be derived for time derivatives of arbitrary order, by taking time derivatives of \Eqref{eq:pde}. 

This 
theorem is mainly a generalization of the corresponding theorem in \cite{Ames:yV5l9m6A} to arbitrary spatial dimensions $n$. However, there is also a small change in \Conditionsref{cond:source1} and \ref{cond:source2}: While in  \cite{Ames:yV5l9m6A},   \Conditionref{cond:source1} must be checked for \textit{all} $w$ in $X_{\delta,\mu,q}$, here this condition must hold necessarily  only for those 
$w$ contained in an arbitrarily small 
closed ball $B_{\delta,\mu,q,s}$. A similar change is made here for \Conditionref{cond:source2}. This change is important in order to encompass  operators $\FreduOp$ for which \Conditionref{cond:source1} is only satisfied if $w$ is bounded in some way. As a particular example we consider  a singular initial value problem, for which $\FreduOp: w\mapsto 1/(1+w)-(1-w)$. If $w\in B_{\delta,\mu,q,s}$ for a choice $\mu>0$ and  for a sufficiently small $s$, then one can show that $\FreduOp[w]\in X_{\delta,2\mu,q}$ and hence \Conditionref{cond:source1} is satisfied. However, \Conditionref{cond:source1} is violated if $s$ is too large, and hence the version of the theorem in \cite{Ames:yV5l9m6A} does not apply. A sketch of the proof of Theorem \ref{theorem:maintheorem} is presented in \Sectionref{sec:proof}.  

We observe that in the hypothesis for Theorem \ref{theorem:maintheorem},  the degree of regularity required for the leading-order matrices $\StLu$, $\SsLu$ and $\NLu$ (specified by $q_0$), and that of the perturbation matrices $\StHu{w}$, $\SsHu{w}$ and $\NHu{w}$, differ slightly from that of the remainder $w$ (given by $q$). The regularity required for the asymptotic data $u_0$ (implicitly specified in \Conditionref{cond:source1})
is also slightly different. These gaps arise in the course of our proof; we note that they arise in particular in working with the higher order energy estimates and the corresponding Cauchy problems for derivatives of $w$ which are needed to control the regularity of solutions. It is not clear if these gaps can be avoided through the use of a different method of proof. In any case, the gaps vanish in the $C^{\infty}$ class of solutions, corresponding to $q=q_0= +\infty$.  

In its applications (see, for instance, \Sectionref{example} below), this theorem often allows one to find an open set of values for the exponent vector $\mu$ for which there exists a unique solution to the singular initial value problem. 
A lower bound for this set\footnote{A number $\Lambda$ is defined to be a lower bound for the allowed values of the vector $\mu$ if each component $\nu^a$ of $\nu$ satisfies the condition $\nu^a(t,x) > \Lambda$ for all $x$ in the domain of $\mu$. A similar definition holds for an upper bound for $\mu$.} can originate in
 \Conditionref{cond:EDM}, while an upper bound is usually determined by \Conditionref{cond:source1}. Both bounds on the set of allowed values for $\mu$ provide 
  useful information on the problem. The upper bound for $\mu$ specifies  the smallest regularity space and, hence, the most precise description of the behavior of $w$ (in the limit $t\searrow 0$), while the lower bound for $\mu$ determines the largest space in
which the solution $u$ is guaranteed to be unique. We note  that this \textit{uniqueness} property must be interpreted with care: under the conditions of our theorem, there is exactly one solution $w$ in the space $X_{\delta,\mu,q}$, although we do not exclude the possibility that
 another solution may exist in a larger space, for example, in $X_{\delta,\widetilde\mu,q}$ with $\widetilde\mu<\mu$. Note that even if a given system does not satisfy our hypothesis above, there exists a systematic method which allows one to
 ``improve" a leading--order term $u_0$; cf.~the discussion of (order-n)-leading-order terms given in \cite{Ames:yV5l9m6A}, or Section~2.2.4 in \cite{Kichenassamy:2007tr}.


\subsection{Example}
\label{example}

It is instructive to discuss an example PDE system which has some of the features of the systems used to model cosmological solutions to Einstein's vacuum equations. We consider the following first-order system in two spatial dimensions (with spatial coordinates $(x, y)$): 
\begin{align}
\label{eq:u1eq}
Du_1 - u_2 =& 0, \\
D u_2 -\lambda u_2- u_1 t \partial_x u_3 - u_1 t \partial_y u_4 =& \, t^{\epsilon} \, u_3 \left( e^{u_1} - u_2 u_4 \right), \\
u_1 Du_3 -u_1u_3-u_1t\partial_x u_2 =& 0, \\
\label{eq:u4eq}
u_1 Du_4 -u_1u_4-u_1t\partial_y u_2 =& 0, 
\end{align}
where $\lambda>0$ and $\epsilon > 0$ are given parameters. This system is derived from the following second--order, quasilinear Euler-Poisson-Darboux (EPD)-type equation
\begin{align}
\label{eq:EPD}
D^2u -\lambda Du - u t^2\left(u_{xx} + u_{yy}\right) = t^{\epsilon} t\partial_x u \left( e^{u} - t^2 \partial_t u \partial_y u \right),
\end{align}
provided we define the new unknowns 
\begin{equation}
\label{stuff}
u_1 := u, \quad u_2 := Du,\quad u_3 := t\partial_x u,\quad u_4 := t \partial_y u. 
\end{equation}
To recover solutions of \Eqref{eq:EPD} from solutions of \Eqref{eq:u1eq}--\eqref{eq:u4eq}, we must impose the third and fourth of the relations in \Eqref{stuff} as constraints on the choice of initial data (for the Cauchy problem) or constraints on the choice of asymptotic data (for the singular initial value problem), and then verify that the evolution preserves these constraints. However, we ignore this issue in this discussion.

Based on considerations discussed in \cite{Beyer:2010tb}, one might guess that a solution of \Eqref{eq:EPD} is likely to behave as $h+t^\lambda g$ near $t\rightarrow 0$, for $h$ and $g$ a pair of $\R$-valued asymptotic data functions. The definition of first-order variables \Eqref{stuff} (and the notion of ODE-leading order terms from \cite{Ames:yV5l9m6A}) suggests the choice
\begin{equation}
\label{eq:leadingorderterm}
u_0=\left(h(x,y)+t^\lambda g(x,y), \lambda t^\lambda g(x,y), t \partial_x h(x,y), t \partial_y h(x,y)\right),
\end{equation}
as a parametrized set of leading order terms. As we determine below, this leading order term, for appropriate choices of $h$ and $g$, together with  \Eqref{eq:u1eq}--\eqref{eq:u4eq}, comprise a quasilinear symmetric
 hyperbolic Fuchsian system (in the sense of~\Defref{def:QSHFsystem}). 
 
 To show this, we substitute $u=u_0+w$ into the system  \Eqref{eq:u1eq}--\eqref{eq:u4eq}, expand in components of $w=(w_1,w_2,w_3,w_4)$, and then comparing with 
  \Eqref{eq:pde} we find that 
\begin{equation}
\label{eq:s0eq}
\Stu{w} = \diag \big(1,1,h + t^\lambda g+w_1, h + t^\lambda g+w_1\big),
\end{equation}
\begin{equation}
S^x(u_0 + w) = \left(
\begin{array}{cccc}
 0 & 0 & 0 & 0 \\
 0 & 0 & -h-t^\lambda g-w_1 & 0 \\
 0 & -h-t^\lambda g-w_1 & 0 & 0 \\
 0 & 0 & 0 & 0
\end{array}
\right),
\end{equation}
\begin{equation}
S^y(u_0 + w) = \left(
\begin{array}{cccc}
 0 & 0 & 0 & 0 \\
 0 & 0 & 0 & -h-t^\lambda g-w_1 \\
 0 & 0 & 0 & 0 \\
 0 & -h-t^\lambda g-w_1 & 0 & 0
\end{array}
\right),
\end{equation}
and
\begin{equation}
\label{eq:nmateq}
N(u_0 + w) = \left(
\begin{array}{cccc}
 0 & -1 & 0 & 0 \\
 0 & -\lambda & 0 & 0 \\
 0 & 0 & -h-t^\lambda g-w_1 & 0 \\
 0 & 0 & 0 & -h-t^\lambda g-w_1
\end{array}
\right),
\end{equation}
where the source term reads 
\[f(u_0+w)=\left(0,t^{\epsilon} (t \partial_x h+w_3) \left( e^{h+t^\lambda g+w_1} - (\lambda t^\lambda g +w_2) (t \partial_y h +w_4) \right),0,0\right)^T.\]
The expression of the matrices $\StZna$, $\SsZna$, and $N_0$ follows directly by setting $t=0$ and $w_1=0$ in the previous expressions.
We then determine  that $\beta_x = \beta_y  =1$. We also see that $h=h(x,y)$ must be positive so that 
\[\StZna= \diag (1,1,h, h)
\] 
is positive definite, and we require that $h\in H^{q_0}(T^2)$ with $q_0=3+q$; we require $q\ge 4$ in order to satisfy \Theoremref{theorem:maintheorem} with $n=2$. Then, $\StZna,\SsZna$,and $N_0$ belong to $H^{q_0}(T^2)$ (see \Defref{def:QSHFsystem}).

We seek solutions with remainder $w \in X_{\delta, \mu, q}$ for an integer $q$ as in \Theoremref{theorem:maintheorem} and for some exponent vector $\mu=(\mu_1,\mu_2,\mu_3,\mu_4)$. 
In order to interpret $w$ as the remainder of a solution with the leading-order term above, we require that $\mu_1,\mu_2>\lambda$ and $\mu_3,\mu_4>1$. The block diagonal conditions, however, force all components of $\mu$ to be the same. We therefore label each component simply by the same symbol $\mu$ and hence require that $\mu(x,y) > \max\{1,\lambda\}$ for all spatial points. The energy dissipation matrix \Eqref{eq:EDM} is
\[M_0=\StZna\,
\begin{pmatrix}
  \mu & -1 & 0 & 0\\
  0 & \mu-\lambda & 0 & 0\\
  0 & 0 & \mu-1 & 0\\
  0 & 0 & 0 & \mu-1
\end{pmatrix},
\]
which is positive definite if $\mu(x,y)>\max\{1,(\lambda+\sqrt{1+\lambda^2})/2\}$ for all spatial points; this is a stronger restriction than $\mu > \max\{1,\lambda\}$.

We now check \Conditionref{cond:source1} of \Theoremref{theorem:maintheorem}, and compute 
\begin{align*}
  &\Fredu{w}\\
  &=\begin{pmatrix}
  0\\
  t^{\epsilon} (t \partial_x h+w_3) \left( e^{h+t^\lambda g+w_1} - (\lambda t^\lambda g +w_2) (t \partial_y h +w_4) \right)
  +t^2(h+w_1)(h_{xx}+h_{yy})\\ 
  \lambda t^{\lambda+1}(h+w_1)g_x\\
  \lambda t^{\lambda+1}(h+w_1)g_y
\end{pmatrix}.
\end{align*}
Here we see that if $h \in H^{q+2}(T^2)$, $g \in H^{q+1}(T^2)$, and $w\in
X_{\delta,\mu,q}$ (for $q\ge 4$ as above) and so long as $\mu(x,y)<\min\{2,1+\epsilon,1+\lambda\}$ at each spatial point $(x,y)\in T^2$, we find
$\Fredu{w}\in X_{\delta,\nu,q}$ for some exponent vector $\nu>\mu$. 

Let us combine the implications of \Conditionref{cond:EDM} and \Conditionref{cond:source1} of \Theoremref{theorem:maintheorem}. For $q\ge 4$, we require that $h\in H^{q+3}(T^2)$ be a positive function, and $g \in H^{q+1}(T^2)$. The upper and lower bounds for the exponent vector $\mu$ can be satisfied if and only if 
\begin{equation}
  \label{eq:conditionlambda}
  0<\lambda<\min\left(\frac {15}8, \frac{3+8\epsilon+4\epsilon^2}{4(1+\epsilon)}\right).
\end{equation}

Finally, \Conditionref{cond:source2} in \Theoremref{theorem:maintheorem} follows immediately from the following
 two lemmas, which were established in Appendix B of \cite{Ames:yV5l9m6A} and readily extend to general $n \geq 1$. 

\begin{lemma} 
Fix any integer $q>n/2$ and let 
  $\phi_1,\phi_2:\R^d\rightarrow\R$ be $q$-times continuously differentiable functions such that $w\mapsto
  \phi_1(w)$ is a map $X_{\delta,\mu,q}\rightarrow X_{\delta,\nu_1,q}$
  and $w\mapsto \phi_2(w)$ is a map $X_{\delta,\mu,q}\rightarrow
  X_{\delta,\nu_2,q}$, for (scalar) exponents $\nu_1$,
  $\nu_2$, and exponent vector $\mu$. Moreover, suppose that for each $s>0$ and for all
  $\bar\delta\in(0,\delta]$, there exist constants
  $C_1,C_2>0$ (which may depend on $s$ but not on $\bar\delta$) such that
  \begin{align*}    
    \|\phi_1[w]-\phi_1[\widetilde w]\|_{\bar\delta,\nu_1,q}
    &\le C_1 \|w-\tilde w\|_{\bar\delta,\mu,q},\\
    \|\phi_2[w]-\phi_2[\widetilde w]\|_{\bar\delta,\nu_2,q}
    &\le C_2 \|w-\tilde w\|_{\bar\delta,\mu,q},
  \end{align*}
  for all $w, \widetilde w\in B_{\bar\delta,\mu,q,s}$.  Now, let
  $\theta:=\phi_1\, \phi_2$.  Then it follows that $w\mapsto \theta(w)$ is a map
  $X_{\delta,\mu,q}\rightarrow X_{\delta,\nu_1+\nu_2,q}$. Moreover,
  for each $s>0$ and for all $\bar\delta\in(0,\delta]$, there exists a
  constant $C>0$ (which in general depend on $s$, but not on $\bar\delta$) such that, 
  for all $w, \widetilde w\in B_{\bar\delta,\mu,q,s}$, 
  \begin{align*}    
    \|\theta[w]-\theta[\widetilde w]\|_{\bar\delta,\nu_1+\nu_2,q}
    \le C\|w-\tilde w\|_{\bar\delta,\mu,q}.
  \end{align*}
\end{lemma}

\begin{lemma} 
  Suppose $q>n/2$. Let $\psi^{(i)}:=\exp\circ\Pi_i$, where
  $\Pi_i:\R^d\rightarrow\R$ is the projection to the $i$th
  component of $d$-vectors. Then $w\mapsto \psi^{(i)}(w)$ is a map
  $X_{\delta,\mu,q}\rightarrow X_{\delta,0,q}$. Moreover, for each
  $s>0$ and for all $\bar\delta\in(0,\delta]$, there exists a constant
  $C>0$ (which in general depends on $s$ but not on $\bar\delta$) such that,   for all $w, \widetilde w\in B_{\bar\delta,\mu,q,s}$, 
  \begin{align*}    
    \|\psi^{(i)}[w]-\psi^{(i)}[\widetilde w]\|_{\bar\delta,0,q}
    \le C \|w-\tilde w\|_{\bar\delta,\mu,q}. 
  \end{align*}
\end{lemma}

\Theoremref{theorem:maintheorem} now yields the following result. 

\begin{theorem}[Existence and uniqueness  for our example system] 
\label{th:EPD}
Let  $h\in H^{q+3}(T^2)$ be a positive function and $g\in H^{q+1}(T^2)$ with $q\ge 4$. 
  The system \Eqsref{eq:u1eq}--\eqref{eq:u4eq} with parameters $\epsilon,\lambda>0$ 
satisfying \Eqref{eq:conditionlambda}
admits a unique solution $u=u_0+w$ with $u_0$ given by \Eqref{eq:leadingorderterm}
and  with $w\in X_{\delta,(\mu,\mu,\mu,\mu),q}$ for $\mu(x,y)\in ( \max\{1,(\lambda+\sqrt{1+\lambda^2})/2\},\min\{2,1+\epsilon,1+\lambda\} )$ for all $(x,y)\in T^2$.
\end{theorem}

Comparing  \Theoremref{theorem:maintheorem} and \Theoremref{th:EPD}, we note that while in the former there is no direct prescription of the regularity of the leading order term in the hypothesis, in the latter the hypothesis requires that $h$ and $g$ (which make up $u_0$, as stated in \Eqref{eq:leadingorderterm}) lie in the specified function spaces. This is consistent, since the regularity requirements imposed on $S^\alpha_0$ and $N_0$ in  \Theoremref{theorem:maintheorem} in fact imply regularity restrictions on $u_0$; we see this explicitly in this example.

A particular problem of \Theoremref{th:EPD} is that we only construct solutions with remainders $w$ in restricted spaces $X_{\delta,(\mu,\mu,\mu,\mu),q}$. In particular, uniqueness only follows in a restricted sense: the full space for the remainder which is compatible with the singular initial value problem associated with the leading-order term \Eqref{eq:leadingorderterm} is given by $\mu_1,\mu_2>\lambda$, and $\mu_3,\mu_4>1$, but here we only obtain uniqueness of solutions in a subspace of this. The full uniqueness result can be obtained by means of (order n)-leading order terms \cite{Ames:yV5l9m6A}.

We remark that if for instance $\lambda\ge 2$, it is in fact reasonable to expect that for the leading-order term \Eqref{eq:leadingorderterm} one cannot guarantee the existence 
of a solution to the singular initial value problem. The heuristic reason is as follows. The leading-order term \Eqref{eq:leadingorderterm} is obtained under the assumption that the source-term and the spatial derivative terms in the equations can be neglected at $t=0$ and that, to leading-order, $(u_1,u_2,u_3,u_4)$ is therefore a solution of the resulting ODE system. However, since for example the second spatial derivative terms in \Eqref{eq:EPD} are $O(t^2)$, which is of the same order as the leading-order term for $\lambda\ge 2$, this assumption breaks down.


\section{Proof of the main theorem}
\label{sec:proof}

\subsection{Preliminary}

The proof of Theorem~\ref{theorem:maintheorem} given in this section closely follows that of the main theorem for systems in one spatial dimension which is presented in \cite{Ames:yV5l9m6A}; therefore, we refer the reader to that paper for the details and only focus here on the
 main steps and point out key differences which occur for space dimension $n>1$.

There are two main steps to the existence proof. First, we  prove an existence theorem for a linear singular initial value problem of Fuchsian form
and, second, we use this result to  formulate a contraction mapping argument for the full quasilinear problem. To prove existence to the linear problem we use a sequence of  Cauchy problems with the initial times $t_j$
characterizing each of them approaching 
$t=0$, and 
we use energy estimates  to show that the sequence of solutions to these Cauchy problems does
converge to a solution to the singular problem. We start by defining the relevant class of linear Fuchsian systems. To simplify the notation for the linear theory, we set $u_0 = 0$ (without loss of generality), so that the solution $u$ coincides with the remainder $w$. 

Which aspects of the proof of Theorem~\ref{theorem:maintheorem} are most affected by the replacement of one space dimension by more than one? Since Sobolev estimates and embeddings depend on dimension, wherever these play a role, dimension matters. This is true for the well-posedness theorems for the Cauchy problem for symmetric hyperbolic systems (see, for example, \cite{Taylor:2011wn}), for handling non-smooth coefficients in the linear singular initial value problem (see \Propref{prop:linearexistence}),
and also for the contraction mapping argument. The first of these affects our formulation and derivation of our energy estimates; see \Lemref{lem:energyestimate} and \Lemref{lem:higherorderenergy}, below. The second requires us to work with products of elements of Sobolev spaces; the resulting dimensional dependence can be seen in the Moser estimates (stated for instance in Proposition~3.7 in Chapter~13 of \cite{Taylor:2011wn}).
The use of Sobolev estimates in carrying out the contraction mapping argument results in the requirement that the regularity index $q$ satisfy the condition $q > n/2 +2$ (consistent with the requirement that $q \ge 3$ for one space dimension).


\subsection{Linear symmetric hyperbolic Fuchsian systems}

As noted,  the starting point for our proof of Theorem \ref{theorem:maintheorem} is a study of linear symmetric hyperbolic Fuchsian systems. We define these now, obtain energy estimates for them, and then use those estimates to prove existence and uniqueness of solutions 
to the singular initial value problem for linear Fuchsian systems.

\begin{definition}[Linear symmetric hyperbolic Fuchsian systems]
\label{def:LSHFS}
Suppose that $\delta$ and $r$ are positive real numbers, $q$ and $q_0$ are non-negative integers,  $\mu:T^n\rightarrow\R^d$ is an exponent vector, and $\zeta:T^n\rightarrow\R^{d \times d}$ is an exponent matrix such that $\zeta>0$.
The system \Eqref{eq:pde} is a \keyword{linear symmetric hyperbolic Fuchsian system} 
if
the following conditions are satisfied:
\begin{enumerate}[label=\textit{(\roman{*})}, ref=(\roman{*})] 
\item \label{cond:lineardecomposition} The matrices $S^0, S^j$, and $N$ are independent of $u$ and can be decomposed as  
\begin{align*}
\St{t,x} =& \StL(x) + \StH{t,x}, \\
\Ss{t,x} = &\RR{1-\beta_j(x)} \left( \SsL(x) + \SsO{t,x} \right),\\
\NN{t,x} = &\NL(x) + \NH{t,x},
\end{align*}
for matrices $\StL(x)$ (positive definite and symmetric), $\SsL(x)$ (symmetric) and $\NL(x)$ in $H^{q_0}$, and for matrices $\StH{t,x}$ (symmetric), $\SsO{t,x}$ (symmetric) and $\NH{t,x}$ in $B_{\delta, \zeta, q, r}$. Here $\beta_j: T^n \to \mathbb
R^{d}$ ($j=1,\ldots,n$)  is a collection of exponent vectors with strictly
positive components.
\item 
The constant $\delta$ is sufficiently small so that $\St{t,x}$ is uniformly positive definite. By
      \keyword{uniformly positive definite} we mean that $\St{t,x}$, given by the decomposition in \Conditionref{cond:lineardecomposition}, is positive definite uniformly\footnote{Since we assume in addition above that $S^0_0$ is positive definite and the perturbation $S^{0}_{1}$ is bounded, this implies that $S^0$ is positive definite pointwise at all $(t,x)\in (0,\delta]\times T^n$ for all $S^0_1\in B_{\delta,\zeta,q,r}$ if $q_0$ and $q$ are sufficiently large (thanks to the Sobolev inequalities).} \textit{at all} $t\in(0,\delta]$ in the $L^2$-sense with respect to $x$, and \textit{for all} $S^0_1\in B_{\delta,\zeta,q,r}$.
\item The source term is linear in the sense that 
\begin{equation}
\FPDEu{w}:=f(t,x,u) =  f_0(t,x) + F_1(t,x)w
\end{equation}
with $f_0 \in X_{\delta, \nu, q}$ and the matrix $F_1$ satisfies
 $\RR{\mu} F_1 \RR{\mu}^{-1} \in B_{\delta, \zeta, q, r}$.
Here $\nu$ is a exponent vector with $\nu>\mu$.
\end{enumerate}
\end{definition}

Both in the linear and in the quasilinear theory, the matrices $S^{\sigma}_1$ and $N_1$ are considered to be perturbations of $S^{\sigma}_0$ and $N_0$. In formulating   \Theoremref{theorem:maintheorem}, we seek estimates that are independent of the particular choice of $S^{\sigma}_1$ and $N_1$. This can be done provided these quantities are bounded; 
hence the introduction of the balls $B_{\delta, \zeta, q, r}$. Throughout the proof we keep careful track of which quantities the constants in various estimates are allowed to depend on. To make this precise, it is useful to have  the following definition.

\begin{definition}
\label{def:uniform}
Suppose that \Eqref{eq:pde} is a {linear symmetric hyperbolic Fuchsian system} for a chosen set of the parameters $\delta, \mu, \zeta, q, q_0$ and $r$. Suppose that a particular estimate (e.g., the energy estimate \Eqref{EnergyEstimate}), involving a collection $\mathcal{C}$ of constants,  holds for solutions of \Eqref{eq:pde} under a certain collection of hypotheses $\mathcal H$. The constants $\mathcal{C}$ are defined to be \keyword{uniform} with respect to the system and the estimate so long as  the following conditions hold: 
\begin{enumerate}[label=\textit{(\roman{*})}, ref=(\roman{*})] 
\item For any choice of $\StOna, \SsOna,N_1,$ and $ F_1$ in the perturbation space $B_{\delta,\zeta,q,r}$ as in \Defref{def:LSHFS} which are compatible with the hypothesis $\mathcal H$, the estimate holds for the same set of constants $\mathcal{C}$.
\item If the estimate holds for a choice of the constants  $\mathcal{C}$ for one particular choice of $\delta$, then for every smaller (positive) choice of $\delta$, the estimate remains true for the same choice of  $\mathcal{C}$.
\end{enumerate}
\end{definition}

As mentioned above, the first step in our proof of \Theoremref{theorem:maintheorem} involves setting up a sequence of Cauchy problems (with the initial times approaching zero), and controlling the solutions of these Cauchy problems using energy estimates. In carrying this out, it is convenient to make the temporary assumption that  the perturbation quantities $S^{\sigma}_1$ and $N_1$ and the source functions $f_0$ and $F_1$ belong to smooth subsets of their respective spaces (cf.~\Defref{def:LSHFS}). Presuming this, we say
that \Eqref{eq:pde} has \keyword{smooth coefficients}. A later step in the proof removes this smoothness assumption. We note that although this assumption -- that \Eqref{eq:pde} has smooth coefficients -- implies that 
the quantities $S^{\sigma}_1, N_1, f_0$ and $F_1$ are  differentiable to all orders, it does not \keyword{not} guarantee that all derivatives have controlled asymptotic behavior. This control holds only for a set of derivatives, as labeled by the relevant function space.

We now define the energy functionals of interest, state the energy estimates, and discuss the proof of these estimates. Generalizing the corresponding notion in \cite{Ames:yV5l9m6A}, we define
\begin{equation}
\label{energies}
E_{\mu,\kappa,\gamma}[v](t):=\frac 12 e^{-\kappa t^\gamma}
\scalarpr{S^0(t,\cdot) \RR{\mu}(t,\cdot) v(t,\cdot)}{\RR{\mu}(t,\cdot)
  v(t,\cdot)}_{L^2(T^n)}
\end{equation}
for any function $v:[t_0,\delta] \times T^n \rightarrow \R^d$ (with
$v(t,\cdot) \in L^2(T^n)$ for each $t \in[t_0, \delta]$). In the above definition $S^0$ is the matrix appearing in \eqref{eq:pde}. It is easily proved that these energy functionals are equivalent to the $L^2(T^n)$-norms with equivalence constants which are uniform in the sense of \Defref{def:uniform}, if $S^0$ can be decomposed as \Conditionref{cond:lineardecomposition} of \Defref{def:LSHFS}. Based on this definition, we have the following energy estimate. 

\begin{lemma}[Fundamental energy estimate]
\label{lem:energyestimate}
Suppose that \Eqref{eq:pde} is a linear symmetric hyperbolic Fuchsian system for the parameters $\delta, \mu,\zeta,q, q_0, r$ 
(according to \Defref{def:LSHFS}),
 has smooth coefficients, and is block diagonal with respect to $\mu$, with $q=0$ and $q_0 > n/2 + 1$. 
Suppose also that the energy dissipation matrix  \Eqref{eq:EDM} is positive definite for all $x\in T^n$
and, in addition, $D\StOna,\partial_i \SsOna \in B_{\delta,\xi,0,s}$ for all $i,j=1,\ldots,n$ for some $s>0$ and some 
exponent matrix $\xi$ with strictly positive entries. Then for any initial data $v_{[t_0]} \in H^{q_0}(T^n)$ specified at some $t_0 \in (0,\delta]$, there exists a unique solution $v$ to the corresponding Cauchy problem, 
and there exist 
positive constants $\kappa, \gamma$ and $C$ such that $v$ satisfies the energy estimate
    \begin{equation}
    \label{EnergyEstimate}
    \sqrt{E_{\mu,\kappa,\gamma}[v](t)}\le \sqrt{E_{\mu,\kappa,\gamma}[v](t)}|_{t=t_0}
    +C\int_{t_0}^t
    s^{-1}\|\RR{\mu}(s,\cdot)
    f_0(s,\cdot)\|_{L^2(T^n)}ds 
 \end{equation}
 for all $t\in [t_0,\delta]$. The constants $C$, $\kappa$, and
 $\gamma$ may be chosen to be uniform\footnote{While the constants $C$, $\kappa$ and
   $\gamma$ here can be chosen to be uniform in the sense of \Defref{def:uniform}, there generally does
   {not} exist a choice which holds for all $\delta, \StZna, \SsZna, N_0, \beta_j, r, s, \zeta, \xi, \mu$ and $\nu$.}.
 In particular, if one replaces $v_{[t_0]}$ specified at $t_0$ by any
 $v_{[t_1]}$ specified at any time $t_1 \in (0, t_0]$, then the energy
 estimate holds for the \emph{same} constants $C$, $\kappa$,
 $\gamma$.
\end{lemma}

The proof of \Lemref{lem:energyestimate} is essentially the same as that for the one-dimensional case~\cite{Ames:yV5l9m6A}. We note that the existence of unique solutions to the $n+1$ dimensional Cauchy problem corresponding to \Eqref{eq:pde} (which  follows from, e.g., Proposition~1.7 in Chapter~16 of \cite{Taylor:2011wn}) plays a key role in this result, and that the inequality for $q_0$ stated in the hypothesis of Lemma \ref{lem:energyestimate} is needed to guarantee such existence.  

We also need to control higher order spatial derivatives of the solutions, for which we establish the following energy estimate.   

\begin{lemma}
\label{lem:higherorderenergy}
Consider a linear symmetric hyperbolic Fuchsian system for the given parameters $\delta, \mu, \zeta, q, q_0$ and $r$,
 which satisfies all of the conditions in 
\Lemref{lem:energyestimate}, except that here we allow for arbitrary integers $q\ge 1$ and $q_0>n/2+1+q$. 
Assume as well that\footnote{We note that the condition $\partial_i \SsOna \in B_{\delta,\xi,0,s}$ ($i,j=1,\ldots,n$) of \Lemref{lem:energyestimate} is now implied by the choice $q\ge 1$.} $D\StOna \in B_{\delta,\xi,0,s}$.
Then there exist positive uniform\footnote{We note however that $C$ and $\rho$  generally depend on $q$.} constants $C, \rho$ such that for all $\epsilon>0$,
the solution $v(t,x)$ of \Lemref{lem:energyestimate} satisfies
 \begin{equation}
    \label{eq:EnergyEstimateHigherOrder}
    \begin{split}
      \| \RR{\mu-\epsilon}(t,\cdot) &v (t,\cdot)\|_{H^q (T^n)}\le 
      C\Bigl(    \|\RR{\mu}(t_0,\cdot) v_{t_0}\|_{H^q (T^n)} \\
      &+\int_{t_0}^t
      s^{-1}\left(\|\RR{\mu}(s,\cdot) f_0(s,\cdot)\|_{H^q (T^n)} +s^\rho \| \RR{\mu} v \|_{H^{q-1}(T^n)} \right) ds\Bigr).
  \end{split}
\end{equation}
The same choice of constants $C$ and $\rho$ can be used for any initial time $t_0\in (0,\delta)$.
\end{lemma}
The inequality for $q_0$ comes again from the condition for well-posedness for the Cauchy problem in $n$ spatial dimensions, but now also from the fact that in deriving the above estimate we take $q$ spatial derivatives of the coefficients.

\begin{proof}
We only discuss the case $q=1$, noting that the argument generalizes to higher values of $q$. 
Let $\partial_i$ be any fixed spatial derivative operator $i = 1,...,n$. As in the case $n=1$, we derive an equation for $\partial_i v$ by acting on both sides of \Eqref{eq:pde} with the derivative operator $\partial_i$. Doing  this for all values of the index $i$, we obtain the following system
\begin{equation}
\label{eq:iderequation}
\widehat\Stna\cdot D\partial v+t\widehat\Ssna\cdot t\partial_j\partial v+\widehat N\cdot\partial v=\widehat f_0+\widehat F_1\cdot\partial v,
\end{equation}
for the ``unknown'' $(n\cdot d)$-vector $\partial v:=(\partial_1 v,\ldots,\partial_n v)^T$, the $(n\cdot d)\times (n\cdot d)$-matrices
\[
  \widehat\Stna:=
  \begin{pmatrix}
    \Stna & 0 & \ldots & 0\\
    0 & \Stna & 0 & \ldots\\
    \vdots & 0 & \ddots &\\
    0& \ldots & & \Stna 
  \end{pmatrix},\,\,
   \widehat\Ssna:=
  \begin{pmatrix}
    \Ssna & 0 & \ldots & 0\\
    0 & \Ssna & 0 & \ldots\\
    \vdots & 0 & \ddots &\\
    0& \ldots & & \Ssna 
  \end{pmatrix},\,\,
   \widehat N:=
  \begin{pmatrix}
    N & 0 & \ldots & 0\\
    0 & N & 0 & \ldots\\
    \vdots & 0 & \ddots &\\
    0& \ldots & & N 
  \end{pmatrix},
\]
the $(n\cdot d)$-vector
\[\widehat f_0=\left(\widehat f_0^{(1)},\ldots,\widehat f_0^{(n)}\right)^T,\]
with components
\[\widehat f_0^{(i)}= \partial_i f_0 + (\partial_iF_1 - \partial_i N)v - (\partial_i\Stna)(\Stna)^{-1}(f_0 + F_1 v - Nv),\]
and the $(n\cdot d)\times (n\cdot d)$-matrix
\[\widehat F_1:=\left(F_1^{(i,j)}\right),\]
with components
\[F_1^{(i,j)}:=F_1\delta_{ij}-t \partial_i\Ssna+t \partial_i\Stna (\Stna)^{-1} \Ssna.\]
All these terms can be handled in essentially the same way as they are in the $n=1$ case in \cite{Ames:yV5l9m6A}, thereby producing the inequality \Eqref{eq:EnergyEstimateHigherOrder}.
\end{proof}

To prove that solutions to the \textit{singular} initial value problem for these linear systems exist, it is  useful to work with weak solutions: We first show that weak solutions exist, and then prove that the weak solutions are differentiable enough to be strong solutions. To define weak solutions for the  linear symmetric hyperbolic Fuchsian system \Eqref{eq:pde}, for $\delta$, $\mu$, $\zeta$, $q$, $q_0$ and $r$, we first define a {test function} $\phi: (0, \delta] \times T^n \to \mathbb R^d$ to be any smooth function with the property that $\phi(t, \cdot) \equiv 0$ for all $t > T \in (0, \delta)$. Then, we define the  operators $\mathcal L$ and $\mathcal F$ acting on $w \in X_{\delta, \mu, 0}$ (similar to \cite{Ames:yV5l9m6A}) by setting 
\begin{equation*}
\begin{split}
\langle \mathcal L[w], \phi \rangle
:=
& -\int_0^\delta \int_{T^n} \, 
\Bigl( \langle \RR{\mu} \Stna w, D\phi \rangle 
+ \sum_{j=1}^n \langle \RR{\mu} \Ssna w, t \partial_j \phi \rangle
- \langle \RR{\mu} N w, \phi \rangle \\
&+ \bigl\langle \RR{\mu} 
	\bigl( 
		\Stna + D \Stna + \RR{\mu}^{-1} D\RR{\mu} \RR{\mu} \Stna
\\
&
+ \sum_{j=1}^n \left( 	\RR{\mu}^{-1} (t \partial_j \RR{\mu}) \Ssna + t \partial_j \Ssna \right)
	\bigr) w, \phi \bigr\rangle 
	\Bigr) dx dt,
 \end{split}
 \end{equation*}
 and
\begin{equation*}
\langle \mathcal F[w], \phi \rangle : = \int_0^\delta \int_{T^n} \langle \RR{\mu} \left( f_0 + F_1 w \right) w, \phi \rangle dx dt.
\end{equation*}
In \cite{Ames:yV5l9m6A} we prove a lemma which shows that these operators are well-defined bounded maps on $X_{\delta, \mu, 0}$; this lemma easily extends to the $n$-dimensional case. We call $w\in X_{\delta, \mu, 0}$ a \keyword{weak solution} of the equation \Eqref{eq:pde} if it satisfies $\langle \mathcal L[w] -\mathcal F[w] , \phi \rangle = 0$ for any test function $\phi$.

Suppose now that \Eqref{eq:pde} is a linear symmetric hyperbolic Fuchsian system for $\delta$, $\mu$, $\zeta$, $q$, $q_0$ and $r$ as in \Defref{def:LSHFS} and suppose that it  is block diagonal with respect to $\mu$ and that the energy dissipation matrix is positive definite. Working with $q=0$ and $q_0 > n/2 + 1$, we recall the construction of the  \keyword{approximate solutions} from \cite{Beyer:2010tb,Ames:yV5l9m6A}: Letting  $\{t_m\}$
be a monotonically decreasing sequence of times  in $(0, \delta]$ with limit zero, we define $v_m: (0, \delta] \times T^n \to \mathbb R^d$ to be the piecewise function which is identically zero on $(0, t_m]$ and which satisfies the Cauchy problem for a linear symmetric hyperbolic Fuchsian system \Defref{def:LSHFS} with zero initial data at $t_m$ on $[t_m,\delta]$. It is shown in detail in \cite{Ames:yV5l9m6A} for the one spatial dimension case that, 
 under appropriate conditions, this sequence of approximate solutions converges to a weak solution of the singular initial value problem (as noted above, we presume here that the leading order term for this problem vanishes). The same argument works here for general $n$. The energy estimate \Lemref{lem:energyestimate} is crucial to this argument. This construction of weak solutions 
gives rise to a solution operator $\mathbb H$ which maps $f_0$ to a weak solution. The  arguments above depend on the system having smooth coefficients, so in particular, the operator $\mathbb H$ to this stage is  only defined for $f_0\in X_{\delta,\nu,0}\cap C^\infty((0,\delta]\times T^n)$. However, since $\mathbb H$ is a linear and bounded map $X_{\delta,\nu,0}\cap C^\infty((0,\delta]\times T^n)\rightarrow X_{\delta,\mu,0}$, it can be extended uniquely as a linear bounded operator to the full space $f_0\in X_{\delta,\nu,0}$. We note that  at this stage of the discussion, we retain the smoothness assumptions of the other coefficients above.

The next step of the argument, which is again discussed in detail in \cite{Ames:yV5l9m6A} for the $n=1$ case,  is to show that for  $q\ge 1$ and $q_0 > n/2 + 1+ q$, the energy estimates of \Lemref{lem:higherorderenergy} 
allow one to verify  higher regularity of the solution. In particular, the solution operator $\mathbb H$ can be shown to be a well-defined 
map $X_{\delta,\nu,q}\rightarrow X_{\delta,\mu,q}$. One can also show that the weak solutions  $w$ are differentiable with respect to $t$ and that $Dw\in X_{\delta,\mu,q-1}$. It then follows that $w$ is  a strong solution of the (linear) system. 

We next proceed to remove the smoothness assumption which has been imposed on $S^{\sigma}_1$, $N_1$ and $F_1$. This step is the first one where the generalization to more than one spatial dimension requires some care, as we discuss in the proof of the following proposition.

\begin{proposition}[Existence of solutions to the linear singular initial value problem]
\label{prop:linearexistence}
Suppose that \Eqref{eq:pde} is a linear symmetric hyperbolic Fuchsian system for $\delta$, $\mu$, $\zeta$, $q$, $q_0$ and $r$ as in \Defref{def:LSHFS} and suppose that it  is block diagonal with respect to $\mu$ for $q_0 > q+ n/2+1$ and $q > n/2+1$.
Suppose that the energy dissipation matrix \Eqref{eq:EDM} is positive definite. Then for all $f_0 \in X_{\delta, \nu, q}$ with $\nu > \mu$ there exists a unique solution $w: (0, \delta] \times T^n \to \mathbb R^d$ to the singular initial value problem with zero leading order term such that $w \in X_{\delta, \mu, q}$ and $Dw \in X_{\delta, \mu, q-1}$. The solution operator $\mathbb H: f_0\mapsto w$ satisfies
\begin{equation}
  \label{eq:estimateH}
  \| \mathbb H[f_0] \|_{\delta, \mu, q} \leq \delta^\rho C \| f_0 \|_{\delta, \nu, q}, 
\end{equation}
for some positive uniform constants $C$, $\rho$.
\end{proposition}
We note that this result also holds for the case $q_0,q=\infty$: If the conditions of this proposition are satisfied for all integers $q_0 > q+ n/2+1$ and $q > n/2+1$, then $w \in X_{\delta, \mu, \infty}$ and $Dw \in X_{\delta, \mu, \infty}$. However, the $q$-parametrized sequence of constants $C$ and $\rho$ occurring in the estimate of the solution operator  may in general be unbounded as $q \rightarrow \infty$.

\begin{proof}
  We restrict attention to the case $F_1\equiv 0$. The basic idea  is to set up a sequence of equations $L_{[i]}[w]=f_0$ defined by a sequence of smooth coefficients $S^{\sigma}_{1[i]}$ and $N_{1[i]}$ which converge to the non-smooth coefficients in the $X_{\delta, \zeta, q}$ norm. For each of the corresponding singular initial value problems, we can apply the (smooth coefficient) arguments discussed above, and we thereby obtain a sequence of solutions $w_{[i]} \in X_{\delta, \mu, q}$, $Dw_{[i]} \in X_{\delta, \mu, q-1}$ to this sequence of singular initial value problems, with   $\|w_{[i]}\|_{\delta, \mu, q} \le C \| f_0 \|_{\delta, \nu, q}$. 
  
We now argue (as in \cite{Ames:yV5l9m6A})  that the sequence $w_{[i]}$ converges in $X_{\delta, \mu, q-1}$. To do this, we note that 
the function $w_{[i]} -w_{[j]}$ is a solution of the system 
\begin{equation}
\label{seqL}
 L_{[i]} [w_{[i]} -w_{[j]}] = - \left(L_{[i]} -L_{[j]}  \right)[w_{[j]}], 
 \end{equation}
where we consider the right-hand side to be a given source term. Now, using the properties of the solution operator corresponding to $L_{[i]} $, we determine that the 
 $\|\cdot\|_{\delta,\mu,q-1}$-norm of the left--hand side of \Eqref{seqL} can be estimated by the $\|\cdot\|_{\delta,\nu,q-1}$-norm of the right--hand side, for some $\nu>\mu$.
To estimate the right--hand side, we use the block diagonal condition and the decompositions of the coefficient matrices which follow from \Defref{def:LSHFS} of a
 linear symmetric hyperbolic Fuchsian system. To estimate the resulting products which consequently appear on the right--hand side, we apply the Moser (Sobolev product) estimates, which require that $q-1>n/2$. 
Then, the fact that the sequence of coefficient matrices is a Cauchy sequence implies that $w_{[i]}$ is a Cauchy sequence in $X_{\delta,\mu,q-1}$. We show that the limit $w\in X_{\delta,\mu,q-1}$ is a weak solution, and in fact a strong solution of the equation with $Dw\in X_{\delta,\mu,q-2}$. 

The remaining step for the existence argument  is to show that we can recover the loss of regularity and in fact verify that $w\in X_{\delta,\mu,q}$ and $Dw\in X_{\delta,\mu,q-1}$. The argument for this is the same as for the $n=1$ case \cite{Ames:yV5l9m6A}.
  
The proof of uniqueness follows the same lines as in \cite{Ames:yV5l9m6A}, by noting that the regularity requirements $q>n/2+1$ and  $q_0>n/2+1+q$ are just sufficient in order to satisfy the hypothesis for $q$ and $q_0$ in the fundamental energy estimate (cf.~\Lemref{lem:energyestimate}).
\end{proof}

\subsection{Quasilinear symmetric hyperbolic Fuchsian systems}

We now have the tools needed to prove \Theoremref{theorem:maintheorem} for general quasilinear symmetric hyperbolic Fuchsian systems.
The idea is to construct the following iteration scheme. We start with some seed function $u_{[0]}=u_0+w_{[0]}$ (here $u_0$ is the leading order term, and $w_{[0]}$ is arbitrary) and we linearize the quasilinear system around $u_{[0]}$. Then we apply our theory for linear equations above and hence, under the hypothesis of \Theoremref{theorem:maintheorem}, this linear system has a solution $u_{[1]}$ of the form $u_{[1]}=u_0+w_{[1]}$. The next step is to linearize the quasilinear system around $u_{[1]}$. Again, we can apply the above theory to this system and hence obtain the solution $u_{[2]}=u_0+w_{[2]}$. We thus obtain a sequence $(w_{[j]}) \in X_{\delta, \mu, q}$ with parameters $\delta, \mu, q$ as specified in the hypothesis of the theorem, which we seek to show converges to the remainder of the solution of the nonlinear equation. 

There 
are several steps to carry out in making  this argument precise (and proving convergence). First, one constructs an operator $\mathbb G(u_0): B_{\delta, \mu, q, s} \to X_{\delta, \mu, q}$ which maps $w_{[i]}$ to $w_{[i+1]}$. If the sequence $(w_{[i]})$ were to leave the closed ball $B_{\delta, \mu, q, s}$, however, then only finitely many sequence elements would be defined. Using \Eqref{eq:estimateH} and the first Lipschitz estimate in \Conditionref{cond:source2} of \Theoremref{theorem:maintheorem},  it can be shown that this can be avoided. In fact, we can choose a sufficiently small $\delta'\in (0,\delta]$ so that $\mathbb G(u_0)$ maps every $w\in B_{\delta', \mu, q, s}$ to an element in $B_{\delta', \mu, q, s}$. Hence, if we restrict the elements $w_{[i]}$ as above to the time interval $(0,\delta']$, then the map $\mathbb G(u_0)$ indeed generates a well-defined infinite sequence in $B_{\delta', \mu, q, s}$.  For every $w_1,w_2\in B_{\delta', \mu, q, s}$, we find a Lipschitz estimate of the form
\[\|\mathbb G(u_0)[w_1]-\mathbb G(u_0)[w_2]\|_{\delta',\mu,q-1}\le C \delta'^\rho\|w_1-w_2\|_{\delta',\mu,q-1},\]
for constants $C,\rho>0$, thanks to the second estimate in \Conditionref{cond:source2} of \Theoremref{theorem:maintheorem}.  The reason for considering the $(q-1)$-norm as opposed to the $q$-norm is that we are required to compare a sequence of systems with different coefficients because the systems are obtained by linearizing around different solutions; recall that we are interested in quasilinear equations here. Just as in the proof of \Propref{prop:linearexistence}, this leads to a loss of one derivative. Therefore, since we apply \Propref{prop:linearexistence} in each step of the iteration for solutions with $q-1$ derivatives, we are led  to a stricter inequality on $q$ which is stated in the hypothesis of \Theoremref{theorem:maintheorem}. We can assume without loss of generality that $\delta'$ is so small so that $C\delta'^\rho<1$. This $\delta'$ is referred to as $\tilde\delta$ in the statement of  \Theoremref{theorem:maintheorem}. It follows that $(w_{[i]})$ is a Cauchy sequence in $X_{\delta',\mu,q-1}$ with limit $w$.  However, we know that the sequence $(w_{[i]})$ is bounded in $X_{\delta',\mu,q}$, and hence we can use duality arguments to show that $w\in X_{\delta',\mu,q}$. It is then left to show that $w$ is in fact differentiable in time and hence a solution of the equation. For this, we note that any solution to the nonlinear singular initial value problem is a fixed point of the operator $\mathbb G(u_0)$. The limit $w$ is the unique fixed point of $G(u_0)$ in $X_{\delta',\mu,q}$, and hence it follows that $w$ is the unique solution of the singular initial value problem.


\section{Concluding remarks}
\label{sec:conclusions}

In our previous work  \cite{Ames:yV5l9m6A}, we have used the results for one space dimension to construct a family of smooth (non-analytic) spacetime solutions  to the $3+1$ Einstein vacuum field equations with $T^2$-symmetry, with these spacetimes all characterized by \emph{asymptotically velocity term dominated} (AVTD) behavior in the neighborhood of the spacetime singularity. We expect that a similar conclusion should hold for $U(1)$-symmetric spacetime solutions and, in order to establish such a result, our generalization to Fuchsian systems with more than one spatial dimension should be crucial. As far as the analytic class is concerned, earlier work by one of the authors with Moncrief and Choquet-Bruhat \cite{Isenberg:2002ku,ChoquetBruhat:2004ix,ChoquetBruhat:2006fc} uses the theorem in \cite{Kichenassamy:1999kg} to show that there are analytic $U(1)$-symmetric solutions of Einstein's equations with AVTD behavior. 

We recall that a solution exhibits AVTD behavior if near the singularity the solution evolves according to a simplified system, specifically an ordinary differential equation (ODE), which can be derived from the full Einstein system by dropping spatial derivatives and keeping time derivative terms, only. The verification of such behavior has proven useful in establishing the validity of the strong cosmic censorship conjecture \cite{Chrusciel:1999dk,Ringstrom:2009ji}, at least within restricted classes of solutions. We believe that this work could play a role in extending the class of solutions for which this conjecture is known to hold.


\end{document}